\documentclass[11pt,superscriptaddress,aps,prd,preprint]{revtex4}

\everymath{\displaystyle}
\usepackage{graphicx}

\usepackage[T1]{fontenc}
\usepackage{amsmath}
\usepackage{amssymb}
\usepackage{graphicx}
\usepackage{slashed}
\usepackage{xcolor}

\newcommand{\bea}{\begin{eqnarray}}
\newcommand{\eea}{\end{eqnarray}}

\begin{document}

\title{Temperature effects for $e^-+e^+\rightarrow \mu^-+\mu^+$ scattering in very special relativity}

\author{A. F. Santos}
\email{alesandroferreira@fisica.ufmt.br}
\affiliation{Instituto de F\'{\i}sica, Universidade Federal de Mato Grosso,\\
78060-900, Cuiab\'{a}, Mato Grosso, Brazil}

\author{Faqir C. Khanna \footnote{Professor Emeritus - Physics Department, Theoretical Physics Institute, University of Alberta\\
Edmonton, Alberta, Canada}}
\email{khannaf@uvic.ca; fkhanna@ualberta.ca}
\affiliation{Department of Physics and Astronomy, University of
Victoria,BC V8P 5C2, Canada}

\begin{abstract}

The electron-positron scattering process is investigated in the context of very special relativity (VSR). This theory assumes that the true symmetry of nature is not the full Lorentz group, but some of its subgroups, such as the subgroups $SIM(2)$ and $HOM(2)$. In this context, the cross-section for electron-positron scattering at finite temperature is calculated. The effects of temperature are introduced  using the Thermo Field Dynamics (TFD) formalism. Our result shows that the cross-section is changed due to both effects, the VSR contributions and temperature effects. An estimated value for the VSR parameter using experimental data available in the literature is discussed.

\end{abstract}

\maketitle

\section{Introduction}

The structure of the standard model, as well as all fundamental physics, is based on Lorentz symmetry. However, the search for a new physics beyond the standard model leads to the possibility that Lorentz symmetry is not an absolute symmetry of nature, but only approximate, valid up to a certain limit of energy. Thus, in recent years, Lorentz invariance and its breakdown have been intensively explored. One motivation, among many others, for such an investigation is the quantization of gravity, the deepest puzzle of contemporary physics. Numerous attempts to construct a consistent theory of quantum gravity lead to the Lorentz violation, such as string theory, loop quantum gravity, non-commutative geometry, among others \cite{Kost, Gambini, Carroll, Jacobson}. Therefore, the search for a fundamental theory that unifies all interactions in nature indicates that Lorentz symmetry may fail at high energy scales. As a consequence, a new physics with different degrees of freedom can be explored. In order to investigate these new possibilities, theories that can accommodate deviations from these symmetries, in the context of standard model particle physics or gravitational phenomena, have been constructed. An example is the standard model extension (SME) \cite{SME1, SME2}. The SME is a theoretical framework that includes standard model, general relativity and all possible operators that break Lorentz symmetry. Another interesting alternative to investigate the Lorentz violation is the very special relativity (VSR) \cite{Cohen1, Cohen2}. 

The basic idea of VSR is that Lorentz symmetry is not a fundamental symmetry of nature, this role is reserved for one of its subgroups preserving the basic elements of special relativity. The $SIM(2)$ and $HOM(2)$ are examples of these subgroups. The subgroup $HOM(2)$ consist of three parameters and is generated by $T_1=K_x+J_y$, $T_2=K_y-J_x$ and $K_z$, where $\vec{J}$ and $\vec{K}$ are the generators of rotations and boosts, respectively. While the subgroup $SIM(2)$ is composed of the $HOM(2)$ group plus the $J_z$ generator. An interesting feature of these subgroups is that they both preserve the direction of a light-like four-vector $n_\mu$, that transform as $n\rightarrow e^\varphi n$ under boost in the z direction, and have no invariant tensor fields. This implies that invariant theories under these subgroups have a preferred direction in Minkowski space-time. Local operators preserving $HOM(2)$ or $SIM(2)$ also preserve Lorentz symmetry. On the other hand, non-local terms can be constructed as ratios of contractions of $n_\mu$ with other kinematic vectors. Although these non-local terms violate Lorentz symmetry, they are invariant under $HOM(2)$ or $SIM(2)$. There are many theoretical and phenomenological studies that investigate interesting aspects of the effects of VSR \cite{Cohen3, Vo, MM, Das, Riv, Bufalo, Bufalo2, Soto, Alf, Bufalo3, Bufalo4, Our, Alfaro, Dimakis}. In this paper, the well-known QED (quantum electrodynamics) process, $e^-+e^+\rightarrow \mu^-+\mu^+$ scattering, in VSR at finite temperature is studied. The Thermo Field Dynamics (TFD) formalism is used to introduce temperature effects.

Two different approaches can be used to introduce the effects of temperature into a quantum field theory. (i) The imaginary time formalism \cite{Matsubara} and (ii) the real time formalism. The latter consist of two approaches: (a) the closed time path formalism \cite{Schwinger} and (b) the TFD formalism \cite{Umezawa1, Umezawa2, Umezawa22, Khanna1, Khanna2, Kbook}. Here the TFD formalism is considered. The main characteristic of this formalism is the duplication of the original Fock space and the use of Bogoliubov transformation. The doubling space is defined as ${\cal S}_T={\cal S}\otimes \tilde{\cal S}$, where ${\cal S}$ and $\tilde{\cal S}$ are the original Fock space and tilde space, respectively. This doubling is defined by the tilde or dual conjugation rules. The Bogoliubov transformation introduces a rotation in the tilde and non-tilde variables, and then temperature effects are implemented in the doubled Fock space. Another important feature of this approach is that the temporal evolution of the system and the effects of temperature can be considered simultaneously.

This paper is organized as follows. In section II, the quantum electrodynamics in VSR is presented. A new interaction Lagrangian due to the VSR modifications is discussed. In section III, TFD formalism is briefly introduced. In section IV, the transition amplitude for the electron-positron scattering at finite temperature is calculated. Then the cross-section in VSR at finite temperature for this process is obtained. A numerical estimate for the VSR parameter is analyzed. In section V, some concluding remarks are presented.

\section{QED in Very Special Relativity}

In this section, the main characteristics of QED in VSR are presented. The $SIM(2)$ VSR-invariant Lagrangian that describes the interaction between fermions and photons is given as
\bea
{\cal L}=-\frac{1}{4}\tilde{F}_{\mu\nu}\tilde{F}^{\mu\nu}+\bar{\psi}\left(i\gamma^\mu \nabla_\mu-m_e\right)\psi-\frac{1}{2\alpha}\left(\tilde{\partial}_\mu A^\mu\right)^2,
\eea
where 
\bea
\tilde{F}_{\mu\nu}=\tilde{\partial}_\mu A_\nu-\tilde{\partial}_\nu A_\mu
\eea
is the field strength defined in terms of the wiggled derivative which is given as
\bea
\tilde{\partial}_\mu=\partial_\mu+\frac{m^2}{2}\frac{n_\mu}{n\cdot \partial}.
\eea
Here, $n_\mu=(1,0,0,1)$ is a light-like four-vector that represents the preferred null direction and the $m$ parameter sets the scale for the VSR effects. It is to be noted that, using the wiggle derivative in the field strength the VSR-Maxwell equation leads to
\bea
\left(\partial^2+m^2\right)A^\nu=0.
\eea
This implies that $A^\nu$ is a massive field. Furthermore, it has been shown that the gauge symmetry associated with the VSR provides massive modes 
as in the standard case, i.e., without changing the number of physical polarization states of the photon \cite{Cheon, Rivelles}. It is important to note that, in this context, the photon mass coming from a term that is gauge invariant, unlike the case where a term of the type $m^2 A^\mu A_\mu$ is not gauge invariant.

In the VSR framework, the minimal coupling between fermions and photons is determined by a new gauge invariant covariant derivative that is given as
\bea
\nabla_\mu\psi=D_\mu\psi+\frac{1}{2}\frac{m^2}{n\cdot D}n_\mu\psi,
\eea
where $D_\mu=\partial_\mu-ieA_\mu$ is the usual covariant derivative. The term $1/(n\cdot D)$ in the covariant derivative has a non-local character that leads to an infinite number of interactions in the coupling $e$. Details about these interactions and Feynman rules have been constructed, see \cite{Dunn}. In order to calculate the cross-section for electron-positron scattering, the most important part of the Lagrangian is the interaction part that is  given as
\bea
{\cal L}_{int}=\bar{\psi}{\cal V}^\mu(p_1,p_2)\psi A_\mu,\label{int}
\eea
where ${\cal V}^\mu(p_1,p_2)$ is the vertex which is defined as
\bea
{\cal V}^\mu(p_1,p_2)= -ie\left[\gamma^\mu+\frac{m^2}{2}\frac{\slashed{n}\,n^\mu}{(n\cdot p_1)(n\cdot p_2)}\right].\label{vertex}
\eea
Observe that in this vertex the contribution due to the VSR has a non-local form. Another important quantity for calculations that will be developed is the photon propagator. Then the gauge propagator with massive pole is given as
\bea
i\Delta^{\mu\nu}(k)=\frac{\eta^{\mu\nu}}{\tilde{k}^2}
\eea
with $\tilde{k}^2=k^2-m^2$. It is important to emphasize that the photon propagates massive modes in a gauge invariant framework without changing its number of physical degree of freedom.

In the next section, the TFD formalism is introduced and then the cross-section at finite temperature for $e^-+e^+\rightarrow \mu^-+\mu^+$ scattering in very special relativity is calculated.

\section{TFD formalism}

Here a brief introduction to TFD formalism is presented. The main motivation for constructing the TFD formalism, which is a real time formalism of quantum field theory at finite temperature, is to obtain that $\langle A \rangle=\langle 0(\beta)| A|0(\beta) \rangle$, i.e., the thermal average of any operator {\cal A} is equal to its temperature dependent vacuum expectation value, with $|0(\beta) \rangle$ being a thermal vacuum. Such a development requires two essential elements: (i) the doubling the degrees of freedom in a Hilbert space and (ii) the Bogoliubov transformation. 

The doubling is defined by the tilde ($^\thicksim$) conjugation rules, that associate each operator in $S$ to two operators in $S_T$, where the expanded space is $S_T=S\otimes \tilde{S}$, with $S$ being the standard Hilbert space and $\tilde{S}$ the tilde (dual) space. The map between the tilde $\tilde{A_i}$ and non-tilde $A_i$ operators is defined by the following tilde (or dual) conjugation rules:
\bea
(A_iA_j)^\thicksim & =& \tilde{A_i}\tilde{A_j}, \quad (\tilde{A_i})^\thicksim = -\xi A_i,\\
(A_i^\dagger)^\thicksim &=& \tilde{A_i}^\dagger, \quad (cA_i+A_j)^\thicksim = c^*\tilde{A_i}+\tilde{A_j},\nonumber
\eea
with $\xi = -1$ for bosons and $\xi = +1$ for fermions.  

The other fundamental ingredient of the TFD formalism is the Bogoliubov transformation that introduces a rotation in the tilde and non-tilde variables, thus the effects of temperature are introduced. The Bogoliubov transformations for fermions, with $c_p^\dagger$ and $c_p$ being creation and annihilation operators respectively, are
\bea
c_p&=&\mathsf{u}(\beta) c_p(\beta) +\mathsf{v}(\beta) \tilde{c}_p^{\dagger }(\beta), \label{f1}\\
c_p^\dagger&=&\mathsf{u}(\beta)c_p^\dagger(\beta)+\mathsf{v}(\beta) \tilde{c}_p(\beta),\label{f2}\\
\tilde{c}_p&=&\mathsf{u}(\beta) \tilde{c}_p(\beta) -\mathsf{v}(\beta) c_p^{\dagger}(\beta),\label{f3} \\
\tilde{c}_p^\dagger&=&\mathsf{u}(\beta)\tilde{c}_p^\dagger(\beta)-\mathsf{v}(\beta)c_p(\beta),\label{f4}
\eea
with $\mathsf{u}(\beta) =\cos \theta(\beta)$ and $\mathsf{v}(\beta) =\sin \theta(\beta)$. The anti-commutation relations for creation and annihilation operators are similar to those at zero temperature
\bea
\left\{c(k, \beta), c^\dagger(p, \beta)\right\}&=&\delta^3(k-p),\nonumber\\
 \left\{\tilde{c}(k, \beta), \tilde{c}^\dagger(p, \beta)\right\}&=&\delta^3(k-p),\label{ComF}
\eea
and other anti-commutation relations are null.

The Bogoliubov transformations for bosons are given as
\bea
a_p&=&\mathsf{u}'(\beta) a_p(\beta) +\mathsf{v}'(\beta) \tilde{a}_p^{\dagger }(\beta), \\
a_p^\dagger&=&\mathsf{u}'(\beta)a_p^\dagger(\beta)+\mathsf{v}'(\beta) \tilde{a}_p(\beta),\\
\tilde{a}_p&=&\mathsf{u}'(\beta) \tilde{a}_p(\beta) +\mathsf{v}'(\beta) a_p^{\dagger}(\beta), \\
\tilde{a}_p^\dagger&=&\mathsf{u}'(\beta)\tilde{a}_p^\dagger(\beta)+\mathsf{v}'(\beta)a_p(\beta),
\eea
where $a_p^\dagger$ and $a_p$ are creation and annihilation operators, respectively. Here, $\mathsf{u}'(\beta) =\cosh \theta(\beta)$, $\mathsf{v}'(\beta) =\sinh \theta(\beta)$ and algebraic rules satisfy the relations
\bea
\left[a(k, \beta), a^\dagger(p, \beta)\right]&=&\delta^3(k-p),\nonumber\\
 \left[\tilde{a}(k, \beta), \tilde{a}^\dagger(p, \beta)\right]&=&\delta^3(k-p),\label{ComB}
\eea
and other commutation relations are null.

In the next section, the cross-section for the electron-positron scattering at finite temperature in VSR is calculated.

\section{Cross section for $e^{-}+e^{+}\rightarrow \mu^-+\mu^+$ scattering at finite temperature}

Here, the scattering process $e^{-}+e^{+}\rightarrow \mu^-+\mu^+$ is considered. The main objective is to analyze how the new vertex changes this process. The Feynman diagrams, that describe this scattering process in VSR, are given as:
\begin{figure}[h]
\includegraphics[scale=0.5]{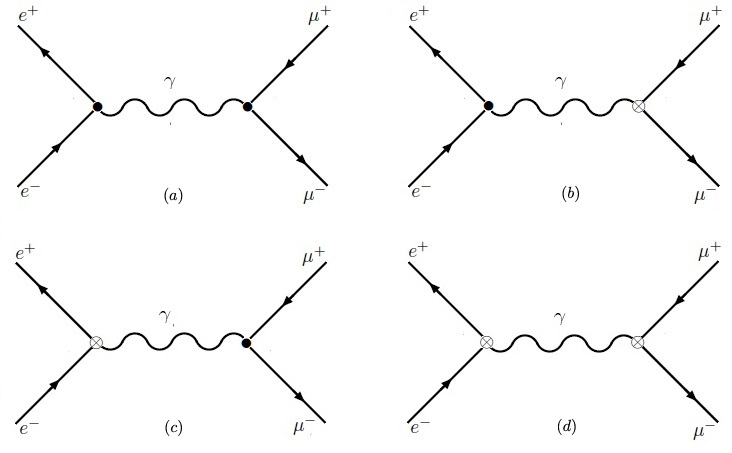}
\caption{Feynman diagrams to electron-positron scattering. The usual QED vertex is denoted by $\bullet$ and the new vertex due to VSR contributions is $\otimes$. Diagram (a) describes the standard QED interaction, while diagrams (b), (c), and (d) describe the new interactions.}
\end{figure}

In the following calculations, the center of mass frame (CM) is considered. Then
\begin{eqnarray}\label{3.11}
&&p_1=(E,p^i),\quad\quad\quad p_3=(E,p'^{i}),\nonumber\\
&&p_2=(E,-p^i),\quad\quad p_4=(E,-p'^{i}),\nonumber\\
&&\kappa=(p_1+p_2)=(\sqrt{s},0),
\end{eqnarray}
with $p_1,\,p_2,\,p_3$ and $p_4$ being the 4-momentum of the electron, positron, muon and anti-muon, respectively and $\sqrt{s}$ is the energy in the center of mass.

In order to calculate the cross-section for the scattering described in FIG.1, the important quantity that must be determined is transition amplitude at finite temperature. It is defined as
\bea
{\cal S}(\beta)=\langle f,\beta| \hat{S}^{(2)}| i,\beta\rangle,
\eea
where 
\bea
| i,\beta\rangle&=& c_{p_1}^\dagger(\beta)d_{p_2}^\dagger(\beta)|0(\beta)\rangle, \nonumber\\
| f,\beta\rangle&=& c_{p_3}^\dagger(\beta)d_{p_4}^\dagger(\beta)|0(\beta)\rangle,
\eea
are the thermal states, with $c_{p_j}^\dagger(\beta)$ and $d_{p_j}^\dagger(\beta)$ being creation operators. In addition, $\hat{S}^{(2)}$ is the second order term of the $\hat{S}$-matrix that is given as
\bea
\hat{S}^{(2)}&=&\frac{(-i)^2}{2!}\int d^4x\, d^4y\, \mathbb{T} \left[ \hat{\cal L}_{int}(x) \hat{\cal L}_{int}(y) \right],
\eea
where $\mathbb{T}$ is the time ordering operator and $\hat{\cal L}_{int}(x)={\cal L}_{int}(x)-\tilde{\cal L}_{int}(x)$ describes the interaction. Then the transition amplitude at finite temperature becomes
\bea
{\cal S}(\beta)&=&\frac{(-i)^2}{2!}\int d^4x\,d^4y\,\langle f,\beta|\mathbb{T}({\cal L}_{int}(x){\cal L}_{int}(y))| i,\beta\rangle.
\eea
It is to be noted that, there is similar equation for the transition amplitude that includes tilde part. 

Using the interaction Lagrangian given by eq. (\ref{int}), we get
\bea
{\cal S}(\beta)&=&-\frac{1}{2}\int d^4x\,d^4y\,\langle f,\beta|\mathbb{T}\left[\bar{\psi}(x){\cal V}^\mu\psi(x)\bar{\psi}(y){\cal V}^\nu\psi(y)A_\mu(x)A_\nu(y)\right]| i,\beta\rangle.
\eea
Considering that the wave function of the fermions field is
\begin{equation}\label{3.3}
\Psi(\mathrm{x})=\int\!\!d\mathrm{p}\Big[c_\mathrm{p}(s)u(\mathrm{p},s)e^{-\imath \mathrm{px}}+d_\mathrm{p}^\dagger(s) v(\mathrm{p},s)e^{\imath \mathrm{px}}\Big],
\end{equation}
with $c_p$ and $d_p$ being annihilation operators for electrons and positrons, respectively with $u(p,s)$ and $v(p,s)$ being Dirac spinors. It is important to note that getting this solution into VSR is not trivial. Non-local terms are introduced and then the canonical quantization must be analyzed. Auxiliary fields are used to performed the quantization. This leads to a non-canonical anti-commutation relation for the fermion field. More details about this study are presented in \cite{Can}. Using the Bogoliubov transformations eq.(\ref{f1}) and eq. (\ref{f2}), the transition amplitude is given as
\bea
{\cal S}(\beta)&=&-\int\frac{d^4p}{(2\pi)^4}\int d^4x\,d^4y\,(\mathsf{u}^2-\mathsf{v}^2)^2\,e^{-ix(p_1-p_3)}e^{-iy(p_2-p_4)}\nonumber\\
&\times&\bar{v}(p_2){\cal V}^\mu(p_1,p_2)u(p_1)\bar{u}(p_3){\cal V}^\nu(p_3,p_4)v(p_4)\langle0(\beta)|\mathbb{T}[A_\mu(x)A_\nu(y)]|0(\beta)\rangle.
\eea

In the TFD approach, the photon propagator at finite temperature is given as
\bea
\langle0(\beta)|\mathbb{T}[A_\mu(x)A_\nu(y)]|0(\beta)\rangle=i\int\frac{d^4\tilde{q}}{(2\pi)^4}e^{-i\tilde{q}(x-y)}\Delta(\tilde{q},\beta)\,\eta_{\mu\nu},
\eea
with
\bea
\Delta(\tilde{q},\beta)=\Delta^{(0)}(\tilde{q})+\Delta^{(\beta)}(\tilde{q})
\eea
where $\Delta^{(0)}(\tilde{q})$ and $\Delta^{(\beta)}(\tilde{q})$ are zero and finite temperature parts respectively, that are defined as
\bea
\Delta^{(0)}(\tilde{q})&=&\frac{1}{\tilde{q}^2}\tau,\nonumber\\
\Delta^{(\beta)}(\tilde{q})&=&-\frac{2\pi i\delta(\tilde{q}^2)}{e^{\beta \tilde{q}_0}-1}\left( \begin{array}{cc}1&e^{\beta \tilde{q}_0/2}\\e^{\beta \tilde{q}_0/2}&1\end{array} \right).
\eea
Here, $\tau=\left( \begin{array}{cc}1 & 0 \\ 0 & -1\end{array} \right)$. More details about photon propagator at finite temperature is given in \cite{first}. By taking $\mathsf{u}(\beta) =\cos \theta(\beta)$ and $\mathsf{v}(\beta) =\sin \theta(\beta)$ leads to $(\mathsf{u}^2-\mathsf{v}^2)^2= \tanh^2\left(\frac{\beta E_{CM}}{2}\right)$. Then the transition amplitude becomes
\bea
{\cal S}(\beta)&=&-i\int\frac{d^4p}{(2\pi)^4}\int d^4x\,d^4y\,\int\frac{d^4\tilde{q}}{(2\pi)^4}\tanh^2\left(\frac{\beta E_{CM}}{2}\right)\,e^{-ix(p_1-p_3+\tilde{q})}e^{-iy(p_2-p_4-\tilde{q})}\nonumber\\
&\times&\bar{v}(p_2){\cal V}^\mu(p_1,p_2)u(p_1)\bar{u}(p_3){\cal V}_\mu(p_3,p_4)v(p_4)\Delta(\tilde{q},\beta).
\eea

Considering the definition of the four-dimensional delta function,
\bea
\int d^4x\,d^4y\,e^{-ix(p_1-p_3+\tilde{q})}e^{-iy(p_2-p_4-\tilde{q})}=\delta^4(p_1-p_3+\tilde{q})\delta^4(p_2-p_4-\tilde{q}),
\eea
and carrying out the $\tilde{q}$ integral leads to
\bea
{\cal S}(\beta)&=&-i\int\frac{d^4p}{(2\pi)^4}\delta^4(p_1+p_2-p_3-p_4)\tanh^2\left(\frac{\beta E_{CM}}{2}\right)\nonumber\\
&\times&\bar{v}(p_2){\cal V}^\mu(p_1,p_2)u(p_1)\bar{u}(p_3){\cal V}_\mu(p_3,p_4)v(p_4)\Delta(\tilde{q},\beta).
\eea
It is interesting note that, the remaining delta function and the $p$ integral express overall four-momentum conservation. By convention, this integral is ignored and the transition amplitude is
\bea
{\cal S}(\beta)&=&-i\tanh^2\left(\frac{\beta E_{CM}}{2}\right)\,\bar{v}(p_2){\cal V}^\mu(p_1,p_2)u(p_1)\bar{u}(p_3){\cal V}_\mu(p_3,p_4)v(p_4)\Delta(\tilde{q},\beta).
\eea
Then the square of the transition amplitude is given as
\bea
\frac{1}{4}\sum_{spin}|i{\cal S}(\beta)|^2&=&\frac{1}{4}\sum_{spin}\tanh^4\left(\frac{\beta E_{CM}}{2}\right)\Delta^2(\tilde{q},\beta)\left[\bar{v}(p_2){\cal V}^\mu(p_1,p_2)u(p_1)\bar{u}(p_1){\cal V}^\nu(p_1,p_2)v(p_2)\right]\nonumber\\
&\times&\left[\bar{u}(p_3){\cal V}^\mu(p_3,p_4)v(p_4)\bar{v}(p_4){\cal V}^\nu(p_3,p_4)u(p_3)\right],\label{square}
\eea
where averaging over the spin of incoming and outgoing particles has been considered and $\Delta^2(\tilde{q},\beta)$ is defined as
\bea
\Delta^2(\tilde{q},\beta)=\frac{1}{\tilde{q}^4}\left[1+\frac{(2\pi\tilde{q}^2)^2\delta^2(\tilde{q}^2)}{(e^{\beta E_{CM}}-1)^2}\right]
\eea
with $\tilde{q}^4=(\tilde{p}_1+\tilde{p}_2)^4$. Using the relation
\bea
\bar{v}(p_2){\cal V}^\mu(p_1,p_2)u(p_1)\bar{u}(p_1){\cal V}^\nu(p_1,p_2)v(p_2)=\mathrm{tr}\left[{\cal V}^\mu(p_1,p_2)u(p_1)\bar{u}(p_1){\cal V}^\nu(p_1,p_2)v(p_2)\bar{v}(p_2)\right]
\eea
and the completeness relations
\bea
\sum_{spin} u(p_1)\bar{u}(p_1)&=&\tilde{\slashed{p}}_1+m_p, \nonumber\\
\sum_{spin} v(p_1)\bar{v}(p_1)&=&\tilde{\slashed{p}}_1-m_p,
\eea
with $\tilde{p}_\mu=p_\mu-\frac{m^2 n_\mu}{2 (n\cdot p)}$ being the wiggle momentum and $m_p$ the particle mass. Hereafter it is assumed that the scattering process is calculated at the high energy limit, which corresponds to $m_p=0$. Then eq. (\ref{square}) becomes
\bea
\frac{1}{4}\sum_{spin}|i{\cal S}(\beta)|^2&=&\frac{{\cal B}(\beta)}{4(\tilde{p}_1+\tilde{p}_2)^4}\mathrm{tr}\left[{\cal V}^\mu(p_1,p_2)\tilde{\slashed{p}_1}{\cal V}^\nu(p_1,p_2)\tilde{\slashed{p}_2}\right]\mathrm{tr}\left[{\cal V}_\mu(p_3,p_4)\tilde{\slashed{p}_4}{\cal V}_\nu(p_3,p_4)\tilde{\slashed{p}_3}\right],
\eea
where
\bea
{\cal B}(\beta)=\tanh^4\left(\frac{\beta E_{CM}}{2}\right)\left[1+\frac{(2\pi\tilde{q}^2)^2\delta^2(\tilde{q}^2)}{(e^{\beta E_{CM}}-1)^2}\right].
\eea

Thus, the differential cross-section for this scattering process is given as
\bea
\frac{d\sigma}{d\Omega}&=&\frac{1}{64\pi^2 s}\frac{1}{4}\sum_{\mathrm{spin}}|i{\cal S(\beta)}|^2,\nonumber\\
&=&\frac{e^4}{64\pi^2 s}\frac{{\cal B}(\beta)\,{\cal J}}{4(\tilde{p}_1+\tilde{p}_2)^4},
\eea
with
\bea
{\cal J}&=&32(\tilde{p}_1\cdot\tilde{p}_4)(\tilde{p}_2\cdot\tilde{p}_3)+32(\tilde{p}_1\cdot\tilde{p}_3)(\tilde{p}_2\cdot\tilde{p}_4)\nonumber\\
&+&16m^2{\cal G}\Bigl[(\tilde{p}_1\cdot\tilde{p}_3)(n\cdot\tilde{p}_2)(n\cdot\tilde{p}_4)+(\tilde{p}_1\cdot\tilde{p}_4)(n\cdot\tilde{p}_2)(n\cdot\tilde{p}_3)+(\tilde{p}_2\cdot\tilde{p}_3)(n\cdot\tilde{p}_1)(n\cdot\tilde{p}_4)\nonumber\\
&+&(\tilde{p}_2\cdot\tilde{p}_4)(n\cdot\tilde{p}_1)(n\cdot\tilde{p}_3)-2(\tilde{p}_3\cdot\tilde{p}_4)(n\cdot\tilde{p}_1)(n\cdot\tilde{p}_2)-2(\tilde{p}_1\cdot\tilde{p}_2)(n\cdot\tilde{p}_3)(n\cdot\tilde{p}_4)\Bigl]
\eea
and 
\bea
{\cal G}&=&\frac{1}{(n\cdot\tilde{p}_1)(n\cdot\tilde{p}_2)}+\frac{1}{(n\cdot\tilde{p}_3)(n\cdot\tilde{p}_4)}.
\eea

From these results and the CM coordinates, the differential cross-section at finite temperature is
\bea
\frac{d\sigma}{d\Omega}&=&\left[\frac{e^4}{128\pi^2\,s}(3+\cos 2\theta)-\frac{e^4\,\Theta^2}{256\pi^2\,s}(19+\cos 2\theta)\right]{\cal B}(\beta),
\eea
where $\Theta\equiv\frac{m}{E}$ is the parameter that controls the VSR effects. The total cross section is obtained by integration
\bea
\sigma=2\pi\, {\cal B}(\beta)\int_0^\pi \left[\frac{e^4}{128\pi^2\,s}(3+\cos 2\theta)-\frac{e^4\,\Theta^2}{256\pi^2\,s}(19+\cos 2\theta)\right]\sin\theta d\theta.
\eea
Performing the integration, the cross-section for electron-positron scattering in VSR at finite temperature is
\bea
\sigma=\sigma_{QED}\left(1-\frac{7}{2}\Theta^2\right){\cal B}(\beta),
\eea
where $\sigma_{QED}=\frac{4\pi\alpha^2}{3s}$ is the usual to QED contribution to the electron-positron scattering. This result shows contributions due to VSR and the finite temperature. It is important to note that the VSR correction term has a similar structure to the QED. There are some experimental results for the cross-section of the electron-positron scattering at zero temperature. Then, from these experimental data, a natural question arises: is it possible to estimate a numerical value for the VSR parameter at the zero temperature limit? In order to answer this question, experimental data \cite{experimental} is used. The VSR correction for this scattering can be assessed using the formula
\bea
\frac{\sigma-\sigma_{QED}}{\sigma_{QED}}=\pm\frac{2s}{\Lambda^2},
\eea
where the center-of-mass energy is $\sqrt{s}=29\,\mathrm{GeV}$ and $\Lambda=170\,\mathrm{GeV}$. Using this data, the estimated value for the VSR parameter is $m\leq 1.9\,\mathrm{GeV}$. It is important to note that, if this parameter is associated with the photon mass, the bound found here is not significant since $m_{photon}<10^{-18}\,\mathrm{eV}$ \cite{mass}. On the other hand, although the effects of VSR alter the usual QED result for electron-positron scattering, experimental data at zero temperature do not establish strong limits on the non-local parameter of VSR. A similar limit for the VSR parameter has been found in the description of the Bhabha scattering \cite{Bufalo3}. Our results show that the temperature effect on the electron-positron scattering may contribute to a new class of constraints on the VSR parameter. Therefore, if the electron-positron scattering is carried out at a finite temperature, a better constraint on the parameters of the model will be obtained.

\section{Conclusion} 

Studies with sufficiently high energies lead to the possibility that a small violation of the Lorentz symmetry may arise. Then, the search for theoretical and phenomenological aspects of the Lorentz violation has increased in recent years. Here, quantum electrodynamics in the VSR Lorentz violating framework is considered. The most interesting feature of the VSR is that the laws of physics are invariant under the (kinematical) subgroups of the Poincaré group, preserving the basic elements of special relativity. In this context, the cross-section for the electron-positron scattering at finite temperature is calculated. The effects of temperature are introduced using the TFD formalism. TFD is a thermal quantum field theory that is composed by two fundamental ingredients: the doubling the degrees of freedom in a Hilbert space and the Bogoliubov transformation. Our result shows that the VSR effects change the usual QED cross-section for this scattering process. However, it is important to note that the VSR effects have the same energy profile as the QED. In addition, the cross-section is affected by the finite temperature. Using experimental data at zero temperature, an estimate for the VSR parameter is obtained. But this binding limit is not significant, since this parameter is compared with the mass of the photon. On the other hand, this limit can be improved if the electron-positron scattering is carried out at high temperatures. Although the deviation predicted in this paper due to the VSR is of theoretical interest, any significant constraints on the parameters of the model will be improved from the experiments performed at non-zero temperature.

\section*{Acknowledgments}

This work by A. F. S. is supported by National Council for Scientific and Technological Develo\-pment - CNPq projects 430194/2018-8 and 313400/2020-2.

\end{document}